\documentclass[11pt]{article}
\usepackage{moriond,epsfig}

\def\be{\begin{equation}}
\def\ee{\end{equation}}
\def\bea{\begin{eqnarray}}
\def\eea{\end{eqnarray}}

\def\sT{\sigma_{\rm T}}
\def\xiacc{\xi_{\rm acc}}
\def\xiload{\xi_{\rm load}}
\def\xigap{\xi_{\rm gap}}
\def\facc{f_{\rm acc}}
\def\Racc{R_{\rm acc}}
\def\Rgap{R_{\rm gap}}
\def\Rload{R_{\rm load}}
\def\Gej{\Gamma_{\rm ej}}
\def\dd{{\rm d}}
\def\bh{\hat{\beta}}
\def\Erad{E_{\rm rad}}
\def\Ed{E_{\rm diss}}
\def\Eej{E_{\rm ej}}
\def\Grel{\Gamma_{\rm rel}}
\def\Rdec{R_{\rm dec}}
\def\mdec{m_{\rm dec}}
\def\tpeak{t_{\rm peak}}
\def\tobs{t_{\rm obs}}

\newbox\grsign \setbox\grsign=\hbox{$>$} \newdimen\grdimen \grdimen=\ht\grsign
\newbox\simlessbox \newbox\simgreatbox \newbox\simpropbox
\setbox\simgreatbox=\hbox{\raise.5ex\hbox{$>$}\llap
     {\lower.5ex\hbox{$\sim$}}}\ht1=\grdimen\dp1=0pt
\setbox\simlessbox=\hbox{\raise.5ex\hbox{$<$}\llap
     {\lower.5ex\hbox{$\sim$}}}\ht2=\grdimen\dp2=0pt
\setbox\simpropbox=\hbox{\raise.5ex\hbox{$\propto$}\llap
     {\lower.5ex\hbox{$\sim$}}}\ht2=\grdimen\dp2=0pt
\def\simgt{\mathrel{\copy\simgreatbox}}

\begin{document}
\vspace*{4cm}
\title{EARLY GRB AFTERGLOWS REVISITED}

\author{ A.M. BELOBORODOV\footnote{Also at Astro-Space Center, 
Profsojuznaja 84/32, Moscow 117810, Russia} }

\address{Canadian Institute for Theoretical Astrophysics, 60 St. George,
M5S 3H8, Canada}

\maketitle\abstracts{
The standard afterglow model neglects the presence of the
GRB prompt radiation ahead of the blast wave. In fact, the
blast wave is dramatically influenced by the leading gamma-ray
front which preaccelerates the ambient medium and loads it with
electron-positron pairs. The front sweeps the medium outward
with a high Lorentz factor and results in a spectacular effect: 
the GRB ejecta moves freely in a cavity behind its own radiation front. 
When the front expands sufficiently and gets diluted, a blast wave 
develops, and it does it differently from the standard model used 
before. The afterglow should initially appear as a steep rise 
of soft emission (from infrared to soft X-rays) at a time comparable 
to the prompt GRB duration and then the emission should quickly 
evolve to a normal X-ray afterglow. This may explain the prompt
optical flash observed in GRB~990123 and allows one to infer 
the ejecta Lorentz factor in this burst: $\Gej\approx 200$. 
The effect of the 
gamma-ray front on the afterglow emission is especially pronounced 
if the GRB has a massive progenitor and the blast wave propagates 
in the progenitor wind. We emphasize the importance of early 
afterglow observations in soft bands, as they will allow one to 
test different progenitor models.
}


\section{Introduction}

The afterglow emission of gamma-ray bursts (GRBs) is believed to come from a 
blast wave driven by a relativistic explosion ejecta into an ambient medium. 
In contrast, the gamma-ray burst itself is emitted early, probably
preceding the development of the blast wave (see [1] for a recent review). 
The $\gamma$-rays decouple from the explosion ejecta, overtake it,
and form a precursor that interacts with the ambient medium ahead of the 
blast wave. The leading radiation front and the ejecta are geometrically 
thin shells, of radius $R$ and thickness $\Delta\ll R$, and they are separated 
by a small distance $l\approx R/\Gamma^2\ll R$ where $\Gamma$ is the ejecta 
Lorentz factor. How does this couple of shells interact with the ambient medium?

In the radiation front, 
two processes take place: Compton scattering and $\gamma-\gamma$ absorption
of the decollimated scattered photons.
As a result the medium is pushed ahead and loaded with $e^\pm$ pairs~[2-4].
The medium dynamics in the front is nonlinear because
the created pairs do more scattering, and the problem is further
complicated by the transfer of scattered radiation through the medium.
Yet an accurate solution can be obtained~[4] and well approximated
by a simple analytical model which is summarized below. 
It turns out that the radiation front changes the medium faced by
the blast wave at radii $R<2\times 10^{16}E_{53}^{1/2}$~cm (where $E$ is the 
``isotropic'' energy of the GRB) and has a dramatic impact on the early 
afterglow. Most of the afterglow energy is likely emitted inside this radius
if the GRB has a massive progenitor. This opens new prospects for testing
progenitor models by studying the early afterglow.

\section{Radiation front}

Let $dE/dS$ be the energy column density of the radiation front [erg/cm$^2$].
Though the explosion does not need to be isotropic, a total ``isotropic''
energy can be formally defined as $E=4\pi R^2(dE/dS)$. When an ambient electron
is overtaken by the radiation front, it scatters energy 
$e_{\rm sc}=\sT(dE/dS)=E\sT/4\pi R^2$. The $e_{\rm sc}$ should be compared to 
the electron rest-mass energy, and this gives a relevant dimensionless parameter,
\begin{equation}
\label{eq:xi}
  \xi=\frac{e_{\rm sc}}{m_ec^2}=65E_{53}R_{16}^{-2}.
\end{equation}
One would like to know what happens with the medium in a radiation front with 
a given $\xi$. In particular, we need to know the velocity $\beta$ acquired by
the medium and the number of loaded pairs per one ambient electron, $f=n_\pm/n_e$.
The answer to this question depends on $\xi$ only. Once one knows
$\beta(\xi)$ and $f(\xi)$ one can substitute Eq.~(\ref{eq:xi}) to find the state
of the medium behind the radiation front at any given $R$. Thus the problem 
reduces to calculation of $\beta(\xi)$ and $f(\xi)$. 
These functions are~[4]
\begin{equation}
  \label{eq:gam}
   \gamma(\xi)=
  \left\{\begin{array}{ll}
    1 & \xi<\xiacc, \\
    (\xi/\xiacc)^3 & \xiacc<\xi<3\xiacc, \\
  3\sqrt{3}(\xi/\xiacc)^{3/2} & \;\; \xi>3\xiacc,\\
  \end{array}\right.
\end{equation}
\begin{equation}
\label{eq:f}
   f(\xi)=
  \left\{\begin{array}{ll}
    \frac{1}{2}[\exp(\xi/\xiload)+\exp(-\xi/\xiload)] & \xi<\xiacc, \\
    (\xi/\xiacc)^2\facc & \xiacc<\xi<3\xiacc, \\
  3(\xi/\xiacc)\facc & \xi>3\xiacc,\\
  \end{array}\right.
\end{equation}
where $\xiload=20-30$, depending on the spectrum of the gamma-rays,
$\xiacc=5\xiload=100-150$, and 
$\facc=[\exp(\xiacc/\xiload)+\exp(-\xiacc/\xiload)]/2=74$.

If $\xi<\xiload$ nothing interesting happens with the medium: it remains 
static and $e^\pm$-free. When the front has $\xi>\xiload$, the runaway $e^\pm$ 
loading occurs. The number of loaded pairs depends exponentially on $\xi$ as 
long as $\xi<\xiacc$.

The front acts as a relativistic accelerator if $\xi>\xiacc$.
The $\xiacc$ is much above unity because the electrons are coupled to the 
ambient ions, and energy $e_{\rm sc}=m_pc^2$ (rather than $m_ec^2$) should be 
scattered per ion to result in relativistic acceleration of the medium. 
On the other hand, $\xiacc=5\xiload$ is 15 times smaller than $m_p/m_e$
because the loaded $e^\pm$ increase the number of scatters per ion.
The relativistic acceleration of the medium is more and more powerful if 
$\xi$ increases above $\xiacc$. At $\xi=\xigap\approx 3\times 10^3$, 
$\gamma$ exceeds the Lorentz factor of the ejecta. It implies that the 
radiation front pushes the medium away from the ejecta and opens a gap.

As the GRB radiation front expands to larger to larger radii, its 
$\xi$-parameter evolves as $\xi\propto R^{-2}$ (Eq.~\ref{eq:xi}).
It starts at very high $\xi$ and then passes through $\xigap$, $\xiacc$,
and $\xiload$ at $\Rgap$, $\Racc$, and $\Rload$, respectively. 
These three characteristic radii define four stages of the explosion:

{\bf I.} $R<\Rgap\approx \Racc/3$. The ejecta moves in a cavity behind its 
own radiation front. The ambient medium surfs ahead with $\gamma>\Gej$.

{\bf II.} $\Rgap<R<\Racc$. The ejecta sweeps the $e^\pm$-rich medium that
has been preaccelerated to $1\ll\gamma<\Gej$.

{\bf III.} $\Racc<R<\Rload\approx 2.3\Racc$. The ejecta sweeps the ``static''
medium ($\gamma\approx 1$) which is still dominated by loaded $e^\pm$.

{\bf IV.} $R>\Rload$. The ejecta sweeps the static pair-free medium.

\noindent
Here we expressed the characteristic radii in terms of $\Racc$ which
is given by
\be
\label{eq:Racc}
   \Racc\approx 10^{16} E_{53}^{1/2}{\rm ~cm}.
\ee


\section{Blast wave}

The blast wave develops at $R>\Rgap$. Technically, its model can be constructed
in the same way as it was done for the standard blast wave in a static medium. 
The only complication is that now the preshock medium is $e^\pm$-rich and
it is moving relativistically.

\subsection{Dynamics: collision with a ``wall''}

The approximate blast wave equations read~[4]
\begin{eqnarray}
\label{eq:dyn1}
  M\frac{\dd\Gamma}{\dd m}=\Gamma^2\bh\gamma(\beta-\bh), \\
  \frac{\dd M}{\dd m}=\eta+(1-\eta)\Gamma\gamma(1-\bh\beta).
\label{eq:dyn2}
\end{eqnarray}
Here $\Gamma=(1-\bh^2)^{-1/2}$ is the Lorentz factor of the blast and 
$\gamma=(1-\beta^2)^{-1/2}$ is the Lorentz factor of the preshock medium.
$M$ is the proper inertial mass of the blast, and $m$ is the swept ambient mass. 
The $m$ is related to radius by $\dd m/\dd R=4\pi R^2 \rho_0$ where $\rho_0(R)$ 
is the medium density ahead of the radiation front.
Eqs.~(\ref{eq:dyn1},\ref{eq:dyn2}) assume that a fraction $\eta$ of the 
dissipated energy is lost to radiation, 
\begin{equation}
\label{eq:Erad}
  \frac{\dd \Erad}{\dd m}=\eta\frac{\dd \Ed}{\dd m}
    =\eta c^2\Gamma\left[\Gamma\gamma(1-\bh\beta)-1\right].
\end{equation}
Two limiting cases are of special interest: $\eta=1$ (radiative blast wave)
and $\eta=0$ (adiabatic blast wave). 

The neglect of the radiation precursor is equivalent to setting $\beta=0$ and
$\gamma=1$ in the above equations. It is interesting to compare the blast wave 
dynamics with and without accounting for the precursor, and we solved the 
equations for both cases to make this comparison. The results are illustrated in 
the upper panel of Fig.~1 which shows how the energy dissipation is distributed 
over radius. In this example, we assumed that the GRB had a Wolf-Rayet 
progenitor with mass loss rate $\dot{M}=2\times 10^{-5}M_\odot$~yr$^{-1}$ and 
wind velocity $w=10^3$~km~s$^{-1}$. Then the ambient mass within a given 
radius $R$ is $m(R)=(\dot{M}/w)R$. 

Without the radiation precursor, one finds that the dissipation rate 
$\dd\Ed/\dd R$ peaks at a ``deceleration'' radius 
$\Rdec\approx 3\times 10^{14}$~cm where $m$ reaches $\mdec=\Eej/\Gej^2c^2$
[$\Eej$ is the initial kinetic energy of the ejecta, and $\Gej=\Gamma(0)$ is 
its initial Lorentz factor]. At $R\sim\Rdec$, $\Gamma$ begins to decline as 
$\Gamma\approx\Gej(m/\mdec)^{-1}$ if $\eta=1$ or 
$\Gamma\approx\Gej(m/\mdec)^{-1/2}$ if $\eta=0$~[5]. 

This standard picture is changed by the $\gamma$-ray precursor.
First of all, the deceleration radius cannot be smaller than 
$\Rgap\approx 3\times 10^{15}E_{53}^{1/2}$~cm. At $R<\Rgap$ the ejecta freely
moves in a cavity cleared by the radiation front. At $R=\Rgap$ the blast wave 
gently begins to sweep the preaccelerated medium with a small relative 
Lorentz factor, $\Grel\approx\Gej/\gamma\approx 1$. With increasing $R>\Rgap$,
$\gamma$ falls off quickly, and it approaches $\gamma=1$ at $R=\Racc$ as 
$\gamma=(R/\Racc)^{-6}$. Thus, after a long delay, the ejecta suddenly 
``learns'' that there is a substantial amount of ambient material on its way 
and hits it with a large $\Grel$. This resembles a collision with a wall and 
results in a sharp peak in the energy dissipation close to the characteristic 
radius $\Racc$, see Fig.~1. 

In our example, the blast wave dissipates 80\% of its energy in the 
preaccelerated zone $R<\Racc$. The relevant dimensionless parameter here 
is~[4]
\be
  D=\frac{2\dot{M} c^2{\Gej}^2}{7w\Eej}\Racc
   \approx 1.8 \left(\frac{\dot{M}}{10^{21}{\rm g/s}}\right)
  \left(\frac{w}{10^{8}{\rm cm/s}}\right)^{-1}
  \left(\frac{E}{10^{53}{\rm erg}}\right)^{1/2}
  \left(\frac{\Eej}{10^{53}{\rm erg}}\right)^{-1}
  \left(\frac{\Gej}{100}\right)^2.
\ee
If $D>1$, the dissipation rate peaks at $R<\Racc$. This peak defines the 
correct deceleration radius, $\Rdec\approx D^{-1/6}\Racc$,
which is very close to $\Racc$ in a wide range of $D$.

\begin{figure}[p]
\psfig{figure=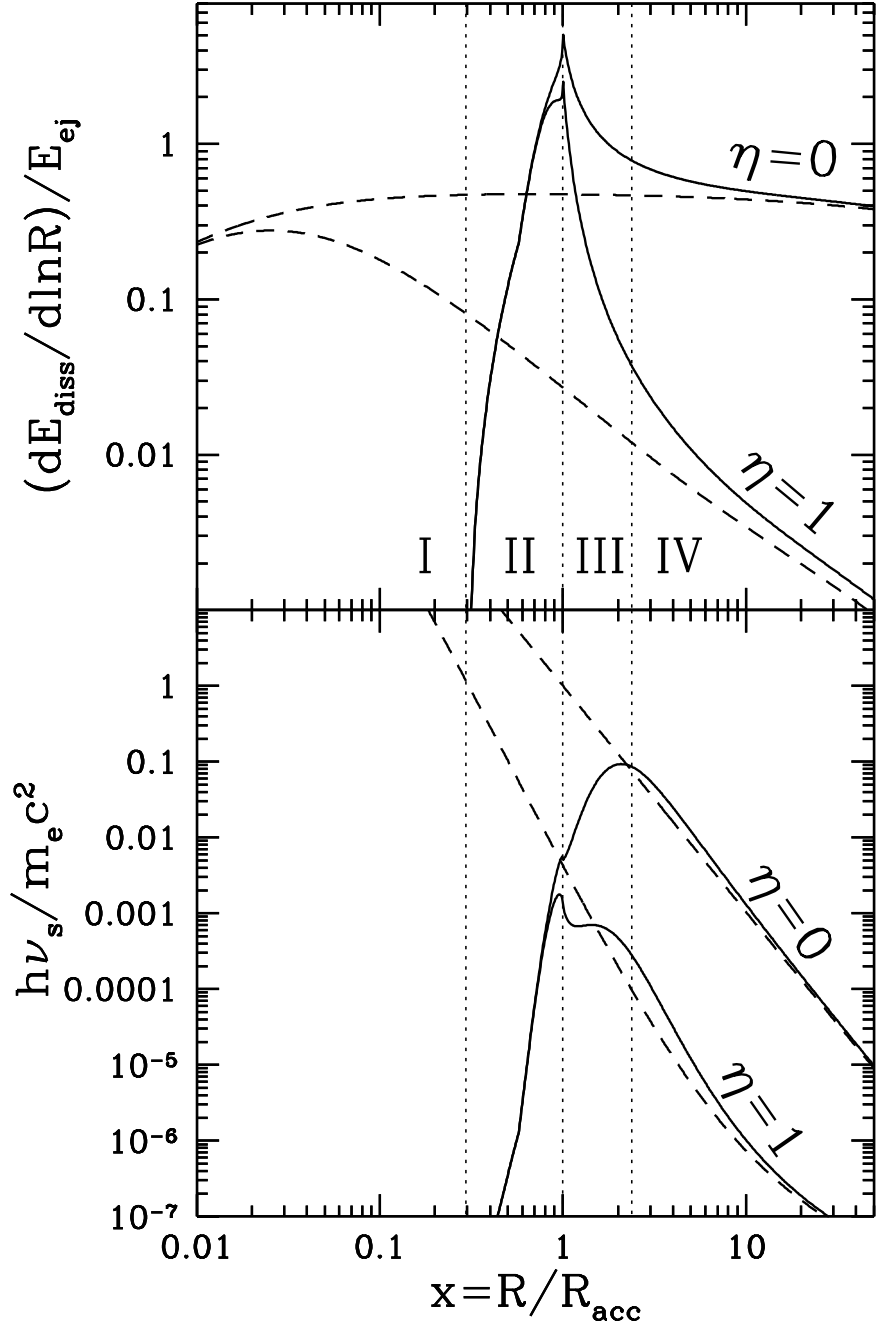,height=18cm}
\caption{
Afterglow from a GRB ejecta decelerating in a wind of a Wolf-Rayet
progenitor with $\dot{M}=2\times 10^{-5}M_\odot$~yr$^{-1}$ and
$w=10^{3}$~km~s$^{-1}$.
The burst is modeled as an impulsive emission of a gamma-ray front
(with isotropic energy $E=10^{53}$~erg) and a thin ejecta shell
with kinetic energy $\Eej=10^{53}$~erg and Lorentz factor $\Gej=200$.
Dashed curves show the prediction of the standard model that neglects
the impact of the radiation front, and solid curves show the actual
behavior. Two extreme cases are displayed in the figure: $\eta=0$ (adiabatic
blast wave) and $\eta=1$ (radiative blast wave). Four zones are marked:
I --- $R<\Rgap$ (the gap is opened),
II --- $\Rgap<R<\Racc$ (the gap is closed and the ejecta sweeps the
relativistically preaccelerated $e^\pm$-loaded ambient medium),
III --- $\Racc<R<\Rload$ ($e^\pm$-loaded ambient medium with
$\gamma\approx 1$), and IV --- $R>\Rload$ (pair-free ambient medium
with $\gamma\approx 1$). Radius is measured in units of
$\Racc\approx 10^{16}$~cm. \hspace{15cm}
{\it Top panel}: dissipation rate. {\it Bottom panel}: synchrotron
peak frequency (assuming $\epsilon_B=0.1$) in units of $m_ec^2/h$.
\label{fig:radish}}
\end{figure}

\subsection{Emission: the optical flash}

One can easily calculate the bolometric light curve produced by the blast 
wave~[4].
It will peak at an observed time $\tpeak$ after the beginning of the burst,
\begin{equation}
\label{eq:tpeak}
  \tpeak\approx\frac{\Racc}{2\Gej^2}\times
  \left\{\begin{array}{ll}
    1 & D<1 \\
    D^{1/6} & D>1
  \end{array}\right\}
 \approx 12\,E_{53}^{1/2}\left(\frac{\Gej}{100}\right)^{-2}{\rm~s}.
\end{equation}
(In the last equality we neglected the weak dependence on $D$).
The predicted peak time depends only on the isotropic equivalent of 
the burst energy, $E$, and the initial Lorentz factor of the ejecta, $\Gej$.
If one manages to observe the peak of an afterglow, find $\tpeak$, 
and also measure the burst redshift (which gives $E$), then one can infer 
$\Gej$ --- the unknown and probably most important parameter of the 
GRB phenomenon. 

The mass of $e^\pm$ pairs ahead of the blast is neglected in the dynamic 
equations because $f=n_\pm/n_e<m_p/m_e$, and the medium mass is dominated 
by the ambient ions.
This is why $f$ did not enter Eqs.~({\ref{eq:dyn1}-\ref{eq:Erad}). 
However, $f\gg 1$ and, along with the preacceleration, it crucially impacts 
the emission spectrum of the blast. One can estimate this effect using the 
simplest synchrotron emission model~[1]. The synchrotron 
spectrum of an electron with Lorentz factor $\gamma_e$ peaks at 
\be
  \nu_s=10^6B\gamma_e^2{\rm ~Hz}.
\ee
where $B$ is the postshock magnetic field. $B$ is usually parametrized
in terms of the equipartition value, so that $B^2/8\pi$ is a fraction 
$\epsilon_B$ of the total energy density. Accounting for the 
jump conditions at the shock front, one can write 
\begin{equation}
 B=c\frac{\Gamma}{\gamma}\sqrt{\frac{32\pi\epsilon_B\rho_0}{\gamma(1-\beta)}}.
\end{equation}
Thus, preacceleration to $\gamma>1$ reduces $B$ by a factor of $\gamma^{-1/2}$. 
The shock energy per ion, $(\Gamma/\gamma)m_pc^2$, is shared by $f$ electrons 
and positrons, and the mean randomized Lorentz factor of the leptons is 
\be
  \gamma_e\approx\frac{\Gamma}{f\gamma}.
\ee
Hence the radiation precursor reduces the synchrotron peak frequency $\nu_s$ 
by the factor of $\gamma^{-5/2}f^{-2}$. This is a big effect at all 
$R<\Rload\approx 2\times 10^{16}E_{53}^{1/2}$~cm. The afterglow should
start as a very soft signal, and it quickly hardens as $\gamma$ and $f$ decrease
steeply with radius. A maximum $\nu_s$ is achieved at $R\simgt\Racc$ and then
the usual decay sets in (Fig.~1, bottom panel). 
 
The initial soft flash should have a broad spectrum and can be observed in the 
optical band. Such a prompt optical flash was detected in GRB~990123~[6], and 
it was suggested to come from a reverse shock in the ejecta~[1]. When the 
$\gamma$-ray precursor is taken into account, the soft flash is naturally 
expected from the forward shock: it appears to be an inevitable result of the 
preacceleration and $e^\pm$-loading of the ambient medium. This mechanism works 
regardless the nature of the GRB ejecta and it also applies to ejecta dominated 
by Poynting flux. 

The optical flash in GRB~990123 has $\tpeak\approx 20$~s (corrected for the
redhsift), and the isotropic $\gamma$-ray energy of the burst is 
$E\approx 3\times 10^{54}$~erg~[7]. Eq.~(\ref{eq:tpeak}) then implies
$\Gej\approx 200$ --- possibly the first measurement of the ejecta Lorentz
factor in a GRB! It does not involve $\epsilon_B$, $\rho_0$, beaming angle,
and other poorly known parameters of the afterglow.

Attempts to detect the prompt optical flash in other bursts did not succeed. 
Instead, quite tight upper limits are obtained at times $\tobs\approx 10$~s 
after the beginning of the GRB. The non-detection can have a simple reason: 
the flash occurs earlier than 10~s, which implies $\Gej>100E_{53}^{1/2}$ 
(see Eq.~\ref{eq:tpeak}). The proposed microsatellite {\em ECLAIRs} [8] 
would be extremely helpful in the study of prompt GRBs as it can provide 
data in soft bands on very short timescales, even much shorter than 10~s.

\section{Concluding remarks}

The initial peak of a GRB afterglow is difficult to observe because it
happens early. In X-rays, it probably overlaps with the prompt GRB.
In soft bands, it is difficult to observe the burst quickly.
Future observations may overcome this technical difficulty. 
{\em Swift} will be able to see optical emission 50~s after the beginning 
of the burst, and the proposed {\em ECLAIRs} can detect flashes at much 
smaller times. A detected peak of the afterglow provides valuable 
information on the ejecta Lorentz factor, and its time profile can give
indications on the nature of the GRB progenitors.

The $\gamma$-ray front plays a crucial role for the early afterglow.
If the internal scenario of the prompt GRB [1] is correct then the 
$\gamma$-rays must clear a cavity in the ambient medium and preclude
any afterglow emission at early times. The afterglow should suddenly 
``switch on'' and rise sharply at $\tobs\sim 10E_{53}^{1/2}(\Gej/100)^{-2}$~s.
The initial flash must be very soft, and it can emit a lot of energy before 
a normal X-ray afterglow sets in, especially in the massive progenitor scenario. 

At the afterglow radii, the ambient medium is optically thin (even after 
$e^\pm$ loading), and the bulk of GRB radiation passes freely through it. 
However, the most energetic $\gamma$-rays may encounter a substantial 
$\gamma-\gamma$ opacity made by the scattered radiation [4].
In the massive progenitor scenario, prompt $\gamma$-rays should be absorbed 
above $\epsilon_{\rm br}=5-50$~MeV that depends on the density of the 
progenitor wind. It produces a spectral break which should be easily detected 
by {\em GLAST}. Additional diagnostics is possible if the scattered 
$\gamma$-rays are directly observed as an echo of the burst~[9].

The optically thin medium scatters only a small portion of the $\gamma$-ray 
energy, much smaller than the energy of the blast wave. In view of this fact, 
the strong dynamical impact of the $\gamma$-ray front on the blast wave might 
seem surprising. The clue to this paradox is the high Lorentz factor of the 
ejecta, $\Gej\gg 1$. In the standard model, a low-mass ambient material 
efficiently decelerates the relativistic ejecta: half of the explosion energy 
$\Eej$ is dissipated when the ejecta sweeps a static $m=\Eej/\Gej^2c^2$. 
If the ejecta sends ahead a radiation precursor, it easily preaccelerates the 
medium to $\gamma\gg 1$ by depositing energy $\gamma mc^2\ll\Eej$. Even at 
$\gamma>\Gej$, the deposited energy is small, while the dynamical impact is 
enormous: $m$ runs away and the ejecta moves freely in the cleared cavity. 
Thus dissipation is delayed until the precursor is diluted sufficiently by 
side expansion.

\section*{Acknowledgments}
This research was supported by NSERC and RFBR grant 00-02-16135.


\end{document}